\documentclass[referee,a4paper,12pt,traditabstract]{swsc} 

\usepackage{graphics,graphicx}
\usepackage[caption=false,font=footnotesize]{subfig}
\usepackage{txfonts}
\usepackage{textcomp}
\usepackage{epstopdf}
\usepackage{lineno}
\usepackage{arydshln}
\usepackage[authoryear,round]{natbib}
\usepackage[backref]{hyperref}
\hypersetup{colorlinks=true,citecolor=cyan,urlcolor=cyan,linkcolor=blue}
\usepackage{url}

\begin{document}


   \title{Ready functions for calculating the Martian radiation environment}

   \authorrunning{Guo et al.}

   \author{Jingnan Guo
          \inst{1}
          \and
          Robert F. Wimmer-Schweingruber
          \inst{1}
          \and
          Manuel Grande
          \inst{2}
          \and
          Zoe Hannah Lee-Payne
          \inst{2}
          \and 
          Daniel Matthi\"{a}
          \inst{3}
          }

   \institute{{Institute of Experimental and Applied Physics (IEAP), University of Kiel, Leibnitzstr. 11, 24118 Kiel, Germany. 
              \email{\href{mailto:guo@physik.uni-kiel.de}{guo@physik.uni-kiel.de}}}
    \and {University of Aberystwyth, Aberystwyth, UK}
    \and German Aerospace Center, Linder H\"{o}he, 51147 K\"{o}ln, Cologne, Germany}
             
 
  \abstract
   {It is extremely important to understand and model the Martian radiation environment in preparation for future human missions to Mars, especially during extreme and elevated conditions such as an intense solar energetic particle (SEP) event. Such events may enhance the radiation level drastically and should be forecasted as soon as possible to prevent severe damage to humans and equipment. 
   Besides, the omnipresent galactic cosmic rays (GCRs) also contribute significantly to the radiation in space and on the surface of Mars and may cause long-term damages to current and future missions. 
   Based on GEANT4 Monte Carlo simulations with the Martian atmospheric and regolith environment setup, we have calculated and obtained some ready-to-go functions which can be used to quickly convert any given SEP or GCR proton/helium ion spectra to the radiation dose on the surface of Mars and also at different depth of the atmosphere. We implement these functions to the {RADMAREE tool} under the Europlanet project which can be easily accessed by the public. } 
   

   \maketitle


\section{Introduction}\label{sec:intro}
The energetic particle environment on the Martian surface is different from that in deep space due to the presence of the Martian atmosphere despite of its rather thin column depth ($\sim$ 20 $\pm$ 5 g/cm$^2$). 
Primary particles passing through the Martian atmosphere may undergo inelastic interactions with the ambient atomic nuclei losing their energies and also creating secondary particles via spallation and fragmentation processes.  
These secondary particles may further interact with the atmosphere as they propagate downwards and even with the Martian regolith and finally result in very complex spectra including both primaries and secondaries at the surface of Mars \citep[e.g.,][]{saganti2002}. 
Much work has been done for calculating the surface exposure rates under different galactic cosmic ray (GCR) conditions and solar energetic particle (SEP) events. 
Several models combining particle transport codes with different GCR and/or SEP spectra have been developed and applied for estimating the radiation exposure on the surface of Mars \citep[e.g.,][]{keating2005model, deangelis2006modeling, mckenna2012characterization, ehresmann2011}. \citet{saganti2004radiation} have mapped the modeled radiation exposure on the Martian surface from GCRs. 
\citet{simonsen1990radiation} and \citet{simonsen1992mars} have calculated the surface dose exposures from both GCRs during solar minimum and maximum conditions as well as some significant SEP events.
\citet{townsend2011estimates} considered the transport of possible Carrington-type SEP events through the Martian atmosphere and also through various hemispherical configurations of aluminum shielding to estimate the resulting organ doses and effective doses of such extreme events.
\citet{norman2014influence} have investigated the influence of dust loading on atmospheric ionizing radiation during solar quiet and SEP events. 
\citet{dartnell2007modelling} have also estimated the effect of surface radiation on the likelihood of survival of microbial life in the Martian soil.

Recently, the radiation assessment detector \citep[RAD,][]{hassler2012} on board the Curiosity rover which is part of the Mars Science Laboratory \citep[MSL,][]{grotzinger2012mars} has been providing the first in situ detection of the radiation environment on the surface of Mars since the landing of MSL on 2012 August 6 \citep{hassler2014}. 
These first measurements provide evaluations of the GCR-induced radiation level inside Gale Crater (with an average column depth of about 22 g/cm$^2$ $\pm$ 25\%) and energetic particle spectra for both charged \citep{ehresmann2014, EHRESMANN20173} and neutral \citep{koehler2014, guo2017neutron} particles. 
The background GCR radiation is not constant and its variation is concurrently driven by the change of the Martian atmosphere as well as the heliospheric activities. The radiation dose rate is anti-correlated with the surface pressure which changes both daily and seasonally \citep{rafkin2014, guo2015modeling, guo2017dependence}. 
Meantime it is also modulated by the varying heliopsheric activities both in the long term \citep{guo2015modeling} and in the short term \citep{guo2018measurements} via e.g., solar coronal mass ejections (CMEs) and their associated shocks. 

During the past six years of measurement, RAD has detected only a few SEP events at the Martian surface and most of them are rather insignificant apart from the 2017 September 10th event \citep{zeitlin2018analysis, ehresmann2018energetic, guo2018modeling} which was the first ground level enhancement (GLE) detected at the surface of two different planets: Earth and Mars.  
SEP events are sporadic, often rather impulsive and could be extremely hazardous especially when the observer is magnetically connected to the acceleration region and injection site at the Sun. 
Therefore it is important to reliably model them in order to give immediate and precise alerts for future human missions at Mars. 
To do so, it is essential to well understand the atmospheric effect on incoming particles. 
There are various particle transport codes such as HZETRN \citep{slaba2016solar, wilson2016}, PHITS \citep{sato2013} and GEANT4/PLANETOCOSMICS \citep{desorgher2006planetocosmics} which can be employed for studying the particle spectra and radiation through the Martian atmosphere. 
We also refer the readers to articles collected in a special issue at Life Sciences in Space Research \citep{HASSLER20171} for more recent studies where several models have been applied to calculate the radiation environment at Mars and the modeled results are compared with data from MSL/RAD. 
When modeling the proton and Helium ion spectra, the GEANT4 model has been shown to be rather reliably matching the measurement \citep{MATTHIA201718}. 


In particular, \citet{guo2018generalized} have developed a generalized approach based on the GEANT4/PLANETOCOSMICS transport code to quickly model the Martian surface radiation level of any given incoming proton/helium ion spectra. 
They have applied it to more than 30 large solar events (measured in situ at Earth) thus providing insights into the possible variety of surface radiation environments that may be induced during large SEP events. 
Depending on the intensity and shape of the original solar particle spectra as well as the distribution of particle types, different SEP events may induce entirely different radiation effects on the surface of Mars.
Primary particles with small energies do not have sufficient energy to reach the ground while the exact atmospheric cutoff energy is a strong function of elevation on Mars. 
Therefore, an intense SEP spectra with moderate high-energy component could be well within biological tolerance seen on the surface of Mars, particularly in low-lying places such as Gale Crater, Hellas Planitia, Valles Marineris, etc., where atmospheric shielding is substantially greater than the global average.
In the current study, we have further developed this approach and calculated ready-to-go functions which can be folded with any given SEP or GCR proton/helium ion spectra to obtain the induced radiation dose on the surface of Mars and also at different depth of the atmosphere. 

\section{Model implementation}\label{sec:model}

\subsection{The PLANETOCOSMICS model}\label{sec:planetocosmics}
GEANT4 {(Version g4.10 has been used in this study)} is a Monte Carlo approach widely used for simulating the interactions of particles as they traverse matter \citep{agostinelli2003}. 
PLANETOCOSMICS is based on GEANT4 with a specific application purpose to simulate particles going through and interacting with planetary atmospheres and magnetic fields \citep{desorgher2006planetocosmics}.
Different settings and features {of the planetary enviroment}, e.g. the composition and depth of the atmosphere and the soil, can be used in the simulations. 
Modeling the radiation environment on the surface of Mars using PLANETOCOSMICS has been carried out in various studies \citep[e.g., ][]{dartnell2007modelling, gronoff2015computation, matthia2016martian, ehresmann2011} and has been validated \citep{matthia2016martian} when compared to energetic charged and neutral particle spectra on the surface of Mars measured by MSL/RAD.

{GEANT4 offers a wide variety of models for handling physical processes of particle interactions at different energy ranges \citep{collaboration2017geant4}. The current study employs the emstandar\_opt4 model for electromagnetic interactions and the QGSP\_BIC\_HP model where QGS stands for the Quark Gluon String model for particles with energy larger than tens of GeV, P represents the Precompound model used for de-excitation process for nucleon-induced reactions below 1-2 MeV and HP switches on the high precision neutron elastic and inelastic scattering model for neutrons below 20 MeV. 
Indeed different physics lists could have been utilized here. 
However, \citet{matthia2016martian} have compared different physics models applied to the Martian environment and the predictions of surface GCR dose rate are agreeing with each other within $\sim$ 10 percent of uncertainty which is generally not larger than the uncertainty of the input particle spectra.}

\subsection{The implementation of the Mars Climate Database}\label{sec:mcd}

The Mars Climate Database (MCD, http://www-mars.lmd.jussieu.fr) offers the possibility to access Martian atmospheric properties, such as temperature, density and composition, for different altitudes, seasons and even the time of the day on Mars.
MCD has been developed using different Martian atmospheric circulation models which are further compared and modified by the observation results from past and current Mars missions \citep{lewis1999climate}. 
Therefore it provides an atmospheric environment sufficiently realistic for the purpose of simulating atmospheric interactions with high energetic particles as the column depth is the most relevant factor here. 
In our atmospheric setup for the PLANETOCOSMICS runs, we use the composition, density and temperature profiles from MCD between altitudes of 250 and 0 km, in steps of 100 m, above the ground at the location of Gale Crater (coordinate: 4.5$^\circ$S, 137.4$^\circ$E,) which is the landing site of MSL's rover Curiosity. 
{The soil is represented by a 100-meter layer of SiO$_2$ with a density of 1.7 g/cm$^3$. }

\subsection{The PLANETOMATRIX approach}\label{sec:planetomatrix}

\begin{figure}[ht!]
\centering
\begin{tabular}{cc}
\subfloat[input: proton, output:  downward proton] { \includegraphics[trim=0 0 80 35,clip, scale=0.5]{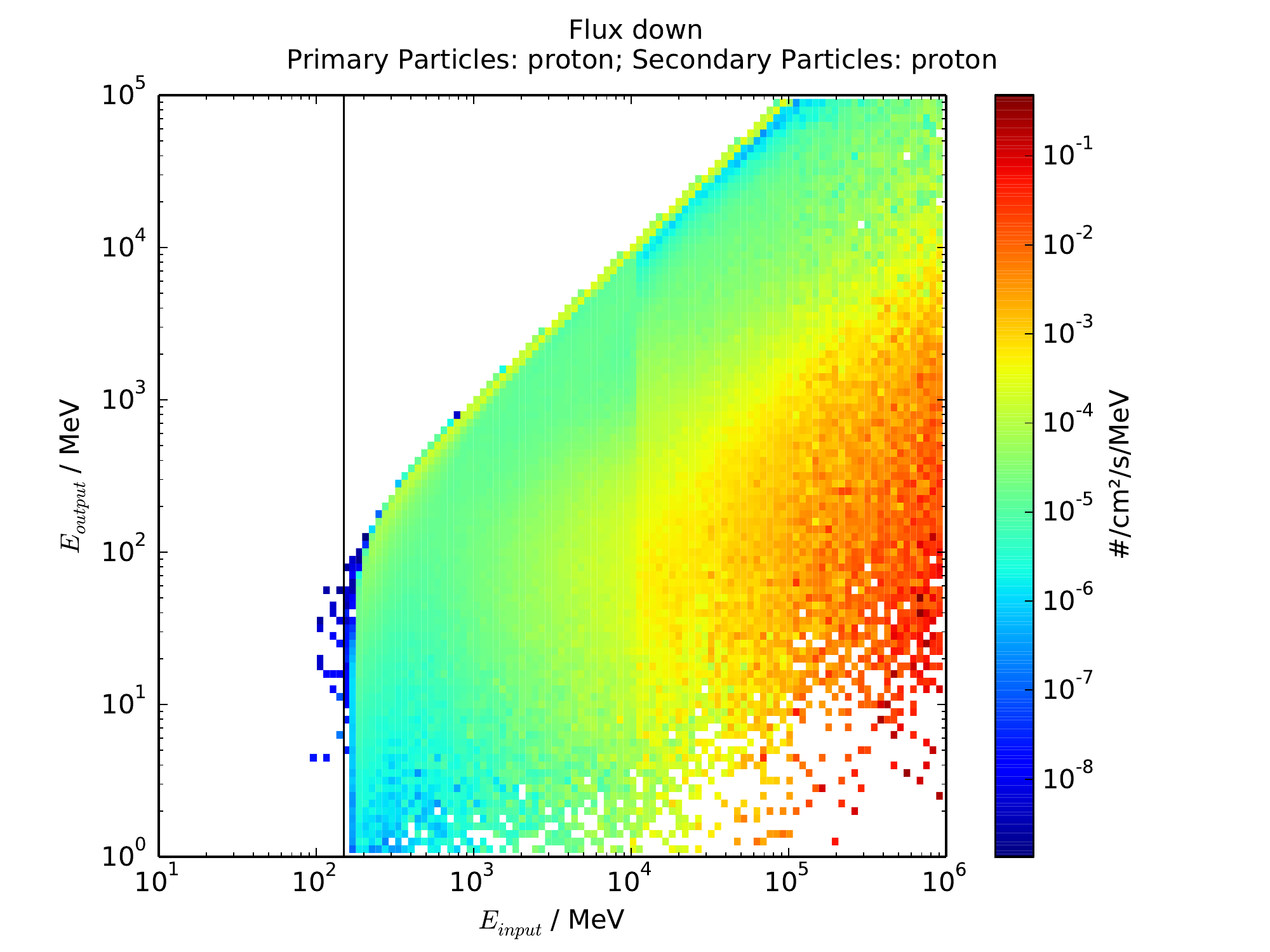}}  & 
\subfloat[input: proton, output:  downward neutron] { \includegraphics[trim=0 0 135 35,clip, scale=0.5]{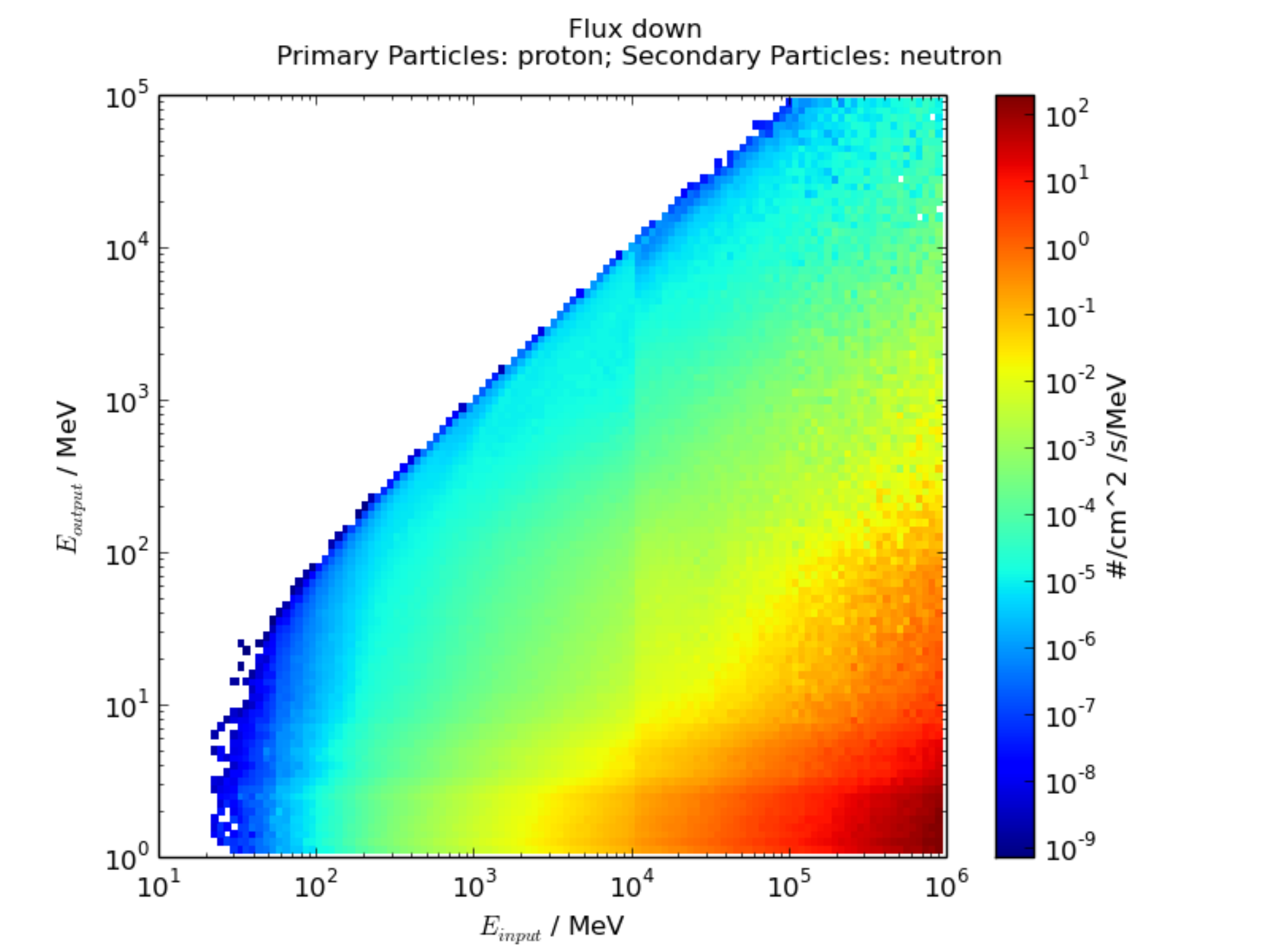} }
\end{tabular}
\caption{2-d histogram of (a) proton-downward proton matrix and (b) proton-downward neutron matrix under a vertical column depth of 20 g/cm$^2$.
X-axis and Y-axis stand for the input and output energies [MeV] respectively. The vertical line in (a) marks the 150 MeV input proton energy.
The probabilities of a proton with certain primary energy producing secondaries with different energies are represented by colors using a logarithmic color distribution. 
Each column in this plot is equivalent to a normalized output spectrum obtained by PLANETOCOSMICS using protons of the corresponding input energy.}\label{fig:PLANETOMATRIX}
\end{figure}

A full PLANETOCOSMICS simulation can be highly time-consuming and in principle needs to be run for each different input spectra. To reduce the computational burden, we have developed an alternative approach which we refer to as the PLANETOMATRIX method \citep{guo2018generalized}. 
Such a matrix $\rm{\bar A_{ij}(E_0, E)}$ represents the "response function" of the Martian atmosphere and it is a statistical description of all interaction process of a particle (of type i) entering the Martian atmosphere with energy $E_0$ resulting in a particle (of type j) with energy $E$ on the Martian surface. 
Although the construction of each matrix is time-consuming, the multiplication of different input spectra $f_{i}(E_0)$ with such a matrix to generate different surface spectra ${F_{j}(E)}$ is very much simplified as shown in Eq. \ref{eq:matrix_multiply}. 
{Specifically speaking, we used 50 million protons simulated with PLANETOCOSMICS which needed approximately 2000 CPU hours. And the time needed for simulating the same number of heavier ions can be much longer due to the production of secondaries. PLANETOMATRIX, on the other hand, only needs linear algebra operation folding the matrix with a given spectra to generate the output spectra and this can be done within a second. }

For input primary particles (particles arriving at the top the Martian atmosphere), we have considered protons and $^4$He ions.
And we included the dominant secondaries such as protons, $^4$He and $^3$He ions, deuterons, tritons, neutrons, gammas, electrons and positrons as output particles on the surface of Mars. 

As particles could also interact with the Martian atmosphere and regolith and produce albedo particles contributing to the upward fluxes, we have also considered the generation of such upward particles in our simulations.  
Since the energy spectra of upward- and downward-traveling particles are dissimilar, we have separately constructed the upward and downward directed matrices for each primary-secondary case as $\rm{\bar A}_{ij}^{\sigma-{up}}$ and $\rm{\bar A}_{ij}^{\sigma-{dn}}$ respectively. The resulting downward and upward spectra of particle type j generated by primary particle type i at the depth of $\sigma$ are respectively \citep{guo2018generalized}:
\begin{eqnarray}\label{eq:matrix_multiply}
{F}_{i,j}^{\sigma-{dn}}(E) = \int \limits_{E_{0}} {\rm{\bar A}}_{ij}^{\sigma-{dn}}(E_0, E) f_{i}(E_0) dE_0; \nonumber \\
{F}_{i,j}^{\sigma-{up}}(E) = \int \limits_{E_{0}} {\rm{\bar A}}_{ij}^{\sigma-{up}}(E_0, E) f_{i}(E_0) dE_0.
\end{eqnarray} 

Panels (a) and (b) of Figure \ref{fig:PLANETOMATRIX} (adapted from and explained in more detail in \citet{guo2018generalized}) show the matrices of primary protons generating downward protons and neutrons respectively. 
The atmospheric depth in this case is about 20 g/cm$^2$. 
It is shown that primary protons with energies less than about 150 MeV, indicated by a vertical line in (a), lack sufficient energy to reach the surface. 
However, these particles could still generate other secondaries such as neutrons which are detected near the surface as shown in Figure \ref{fig:PLANETOMATRIX}b. 

\subsection{The ready-to-go functions for the Martian radiation level}\label{sec:functions}

Radiation dose rate and dose equivalent rate (or gray-equivalent rate for SEPs) are the key quantity for evaluating the energetic particle environment and their potential biological effects for future crewed missions to deep space and to Mars. 
Both charged and neutral particles deposit energy while going through target materials such as skin, bones and internal organs. 
Dose is defined as the energy deposited by radiation per unit mass of the material with a unit of J/kg (or Gy). 
The material throughout this study is considered to be water as it is a good representative for tissue. 
{For a better understanding of the concepts of radiation dose and how to calculate it, please refer to the Appendix of this article.}

The radiation dose rate for either the downward- or the upward-directed particles (of different types j) induced by certain primary particle (of type i) can be calculated following the logic:  
\begin{eqnarray}\label{eq:dose_dn}
D_i^{\sigma-{dn}}(E_0) &=& \sum\limits_{j} {\int \limits_{}^{dn} d\Omega} {\int \limits_{E_{min} }^{E_{max}} F_{i,j}^{\sigma-{dn}}(E) \cdot \lambda_j(E) dE}/{\rho}  \nonumber \\
&=& 2\pi \sum\limits_{j} {\int \limits_{E_{min}}^{E_{max}}
\int \limits_{E_{0}-\Delta E_0/2}^{{E_{0}+\Delta E_0/2}} {\rm{\bar A}}_{ij}^{\sigma-{dn}}(E_0, E) f_{i}(E_0) dE_0 \cdot \lambda_j(E) dE}/{\rho},
\end{eqnarray}
and 
\begin{eqnarray}\label{eq:dose_up}
D_i^{\sigma-{up}}(E_0) &=& \sum\limits_{j} {\int \limits_{}^{up} d\Omega} {\int \limits_{E_{min} }^{E_{max}} F_{i,j}^{\sigma-{up}}(E) \cdot \lambda_j(E) dE}/{\rho}  \nonumber \\
&=& 2\pi \sum\limits_{j} {\int \limits_{E_{min}}^{E_{max}} \int \limits_{E_{0}-\Delta E_0/2}^{{E_{0}+\Delta E_0/2}} {\rm{\bar A}}_{ij}^{\sigma-{up}}(E_0, E) f_{i}(E_0) dE_0 \cdot \lambda_j(E) dE}/{\rho},
\end{eqnarray}
where $f_i(E_0)$ is the primary particle (type i) spectrum (e.g., in the unit of counts/MeV/sec/cm$^2$/sr) and $F_{i,j}^{\sigma-{dn}}(E)$ and $F_{i,j}^{\sigma-{up}}(E)$ are the induced secondary particle spectrum (type j) at certain atmospheric depth $\sigma$ in downward and upward directions respectively. 
$E_{min}$ and $E_{max}$ are the minimum and maximum energies of particles used for the dose calculations. They are set to be 1 and $10^6$ MeV respectively as they are the lower and upper limits of energy bins used for storing output particle histograms.  
$\rho$ is the mass density of the target material.

The energy transfer process of particles in material is expressed as a yield factor $\lambda_j(E)$ which represents the deposited energy per unit length $d\epsilon/dx$ (also called the linear energy transfer (LET)) as a function of incoming particle energy $E$. 
$\lambda_j(E)$ depends on particle species $j$ and the target material. It can be readily estimated using either the Bethe-Bloch equation \citep{bethe1932bremsformel} (for charged particle ionization energy loss in an infinite volume) or with more sophisticated Monte Carlo models such as GEANT4 \citep{matthia2016martian} accounting for the probability distribution of $\epsilon$ in finite volumes as used in this study.
Dose is then the energy loss per unit mass of the material which is calculated as LET divided by the material mass density: $d\epsilon/dx/\rho$.

The resulting $D^{\sigma-{dn}}(E_0)$ and $D^{\sigma-{up}}(E_0)$ in Equations \ref{eq:dose_dn} and \ref{eq:dose_up} are the corresponding dose integrated over the downward or upward directions ($2\pi$ solid angle in each case) respectively. 
For given primary spectra $f_i(E_0)$, dose is summed over all the secondary particle species generated by the primary particles. Taken into account of the flux variation rate in time, dose rate can also be estimated and is often converted to \textmu Gy/day for the radiation level in space. 
Similarly, dose equivalent rate, in \textmu Sv/day, can be calculated as explained later.
The final dose rate at certain atmospheric depth $\sigma$ induced by primary particle flux $f_i$ at energy $E_0$ (with an energy interval of $\Delta E_0$) is simply the sum of the upward and downward dose rates.  

As described in Section \ref{sec:planetomatrix}, we have constructed the upward and downward directed matrices for each primary-secondary pair in the setup of the Martian atmospheric and regolith environment. 
Therefore the dose rate in the above equations \ref{eq:dose_dn} and \ref{eq:dose_up} is purely dependent on $f_i(E_0)$ which is the primary particle spectra as a function of energy $E_0$. 
Therefore we have further developed the PLANETOMATRIX approach and calculated the resulting integrated dose rate described in equations \ref{eq:dose_dn} and \ref{eq:dose_up} induced by a primary particle flux of [1 particle/MeV/sr/cm$^2$/s] in the energy interval from $E_0-\Delta E_0/2$ to $E_0+\Delta E_0/2$ where $\Delta E_0$ is the bin width of the input particle energy in ${\rm{\bar A}}_{ij}(E_0, E)$. 
Such calculations have been repeated for each primary energy bin of the matrices in the range of 20 to 10$^5$ MeV. 
The resulting dose functions depend on the primary particle energy (either protons or helium ions) and are plotted in Figs. \ref{fig:dosefunctions_surf} and \ref{fig:dosefunctions_827Pa}. 
The shown downward and upward functions have been normalized to the input spectra and thus have units of [\textmu Gy/day]/[particles/sr/cm$^2$sec].  
They are mathematically described as $D^{\sigma-{dn}}(E_0)/(f(E_0) \cdot \Delta E_0)$ and $ D^{\sigma-{up}}(E_0)/(f(E_0) \cdot \Delta E_0)$. 
These ready-to-go functions can be folded with any given SEP or GCR proton/helium ion spectra $f(E_0)$ to obtain the induced radiation dose rate (upward, downward and all) on the surface of Mars and also at different depth of the atmosphere. 

In terms of biological effectiveness associated with radiation exposures on human beings, the dose equivalent (in units of Sv) is often more referred to for evaluating the deep space exploration risks \citep{sievert1960}. 
It depends on a quality factor $\rm{Q}$ \citep{icrp60} which is a function of LET. 
Folding $\lambda_j(E)$ with $\rm{Q}(LET)$ into Eqs. \ref{eq:dose_dn} and \ref{eq:dose_up}, one can calculate dose equivalent rate similar to dose rate. Therefore we have also obtained the ready-to-go functions for the dose equivalent rate as a function of primary particle energy as shown in Figure \ref{fig:doseEqfunctions_827Pa}. 

\section{Results and discussions}\label{sec:results}

\begin{figure}[ht!]
\centering
\begin{tabular}{cc}
\subfloat{ \includegraphics[trim=15 20 0 20,clip, scale=0.48]{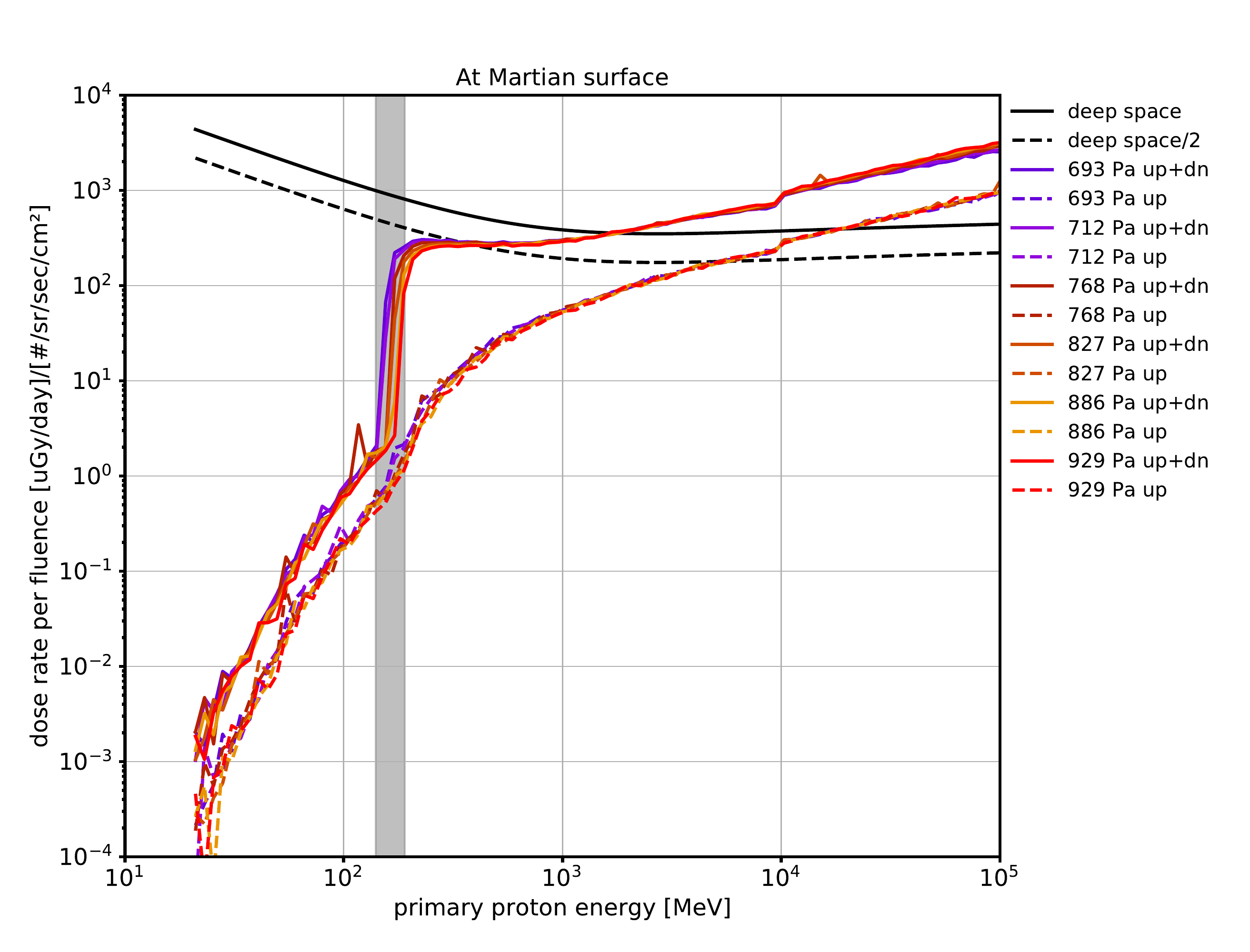}} \\
\subfloat{ \includegraphics[trim=15 20 0 20,clip, scale=0.48]{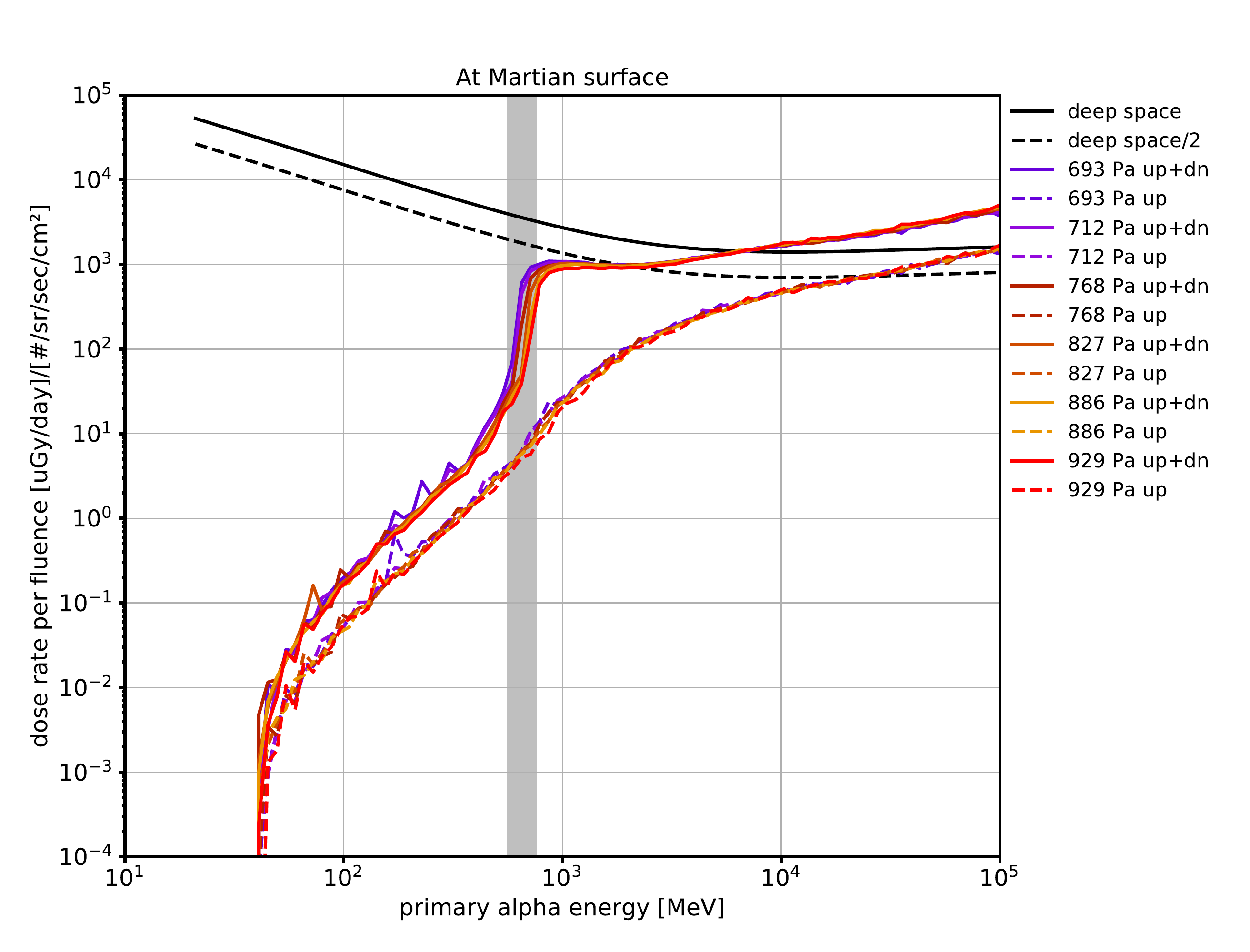}}
\end{tabular}
\caption{Dose rate in water per primary flux function for protons (top) and $^4$He (bottom) on Mars under different surface pressures. Solid lines are the total dose rate while dashed lines represent only the upward secondary contribution. Black solid lines mark the deep space dose per flux while black dashed lines represent half of the deep space case. The gray shaded area marks range (140-190 MeV/nuc) of the atmospheric cutoff energy under different surface pressures. 
}\label{fig:dosefunctions_surf}
\end{figure}
\begin{figure}[ht!]
	\centering
	\begin{tabular}{cc}
		\subfloat{ \includegraphics[trim=15 20 20 20,clip, scale=0.48]{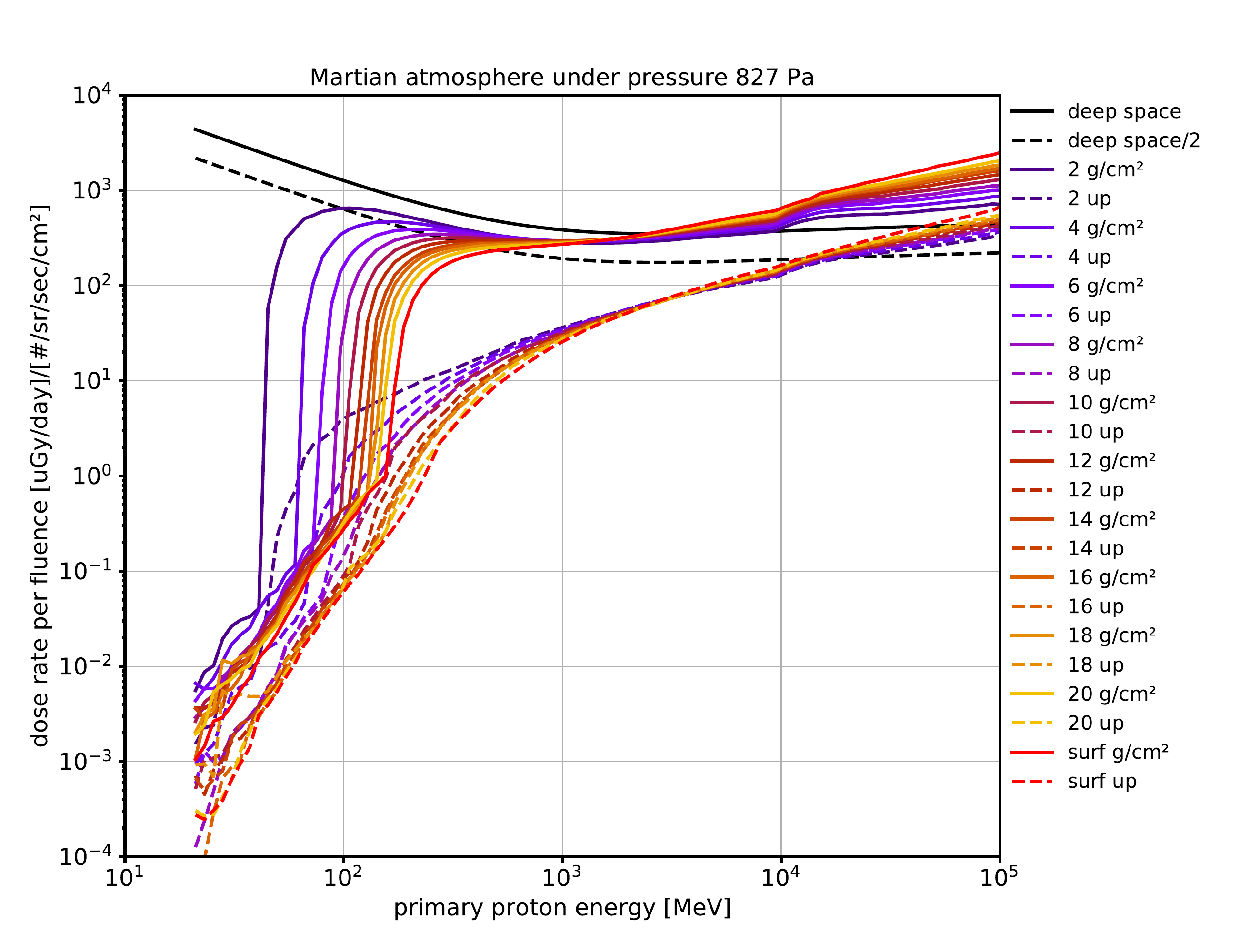} }\\
		\subfloat{ \includegraphics[trim=15 20 20 20,clip, scale=0.48]{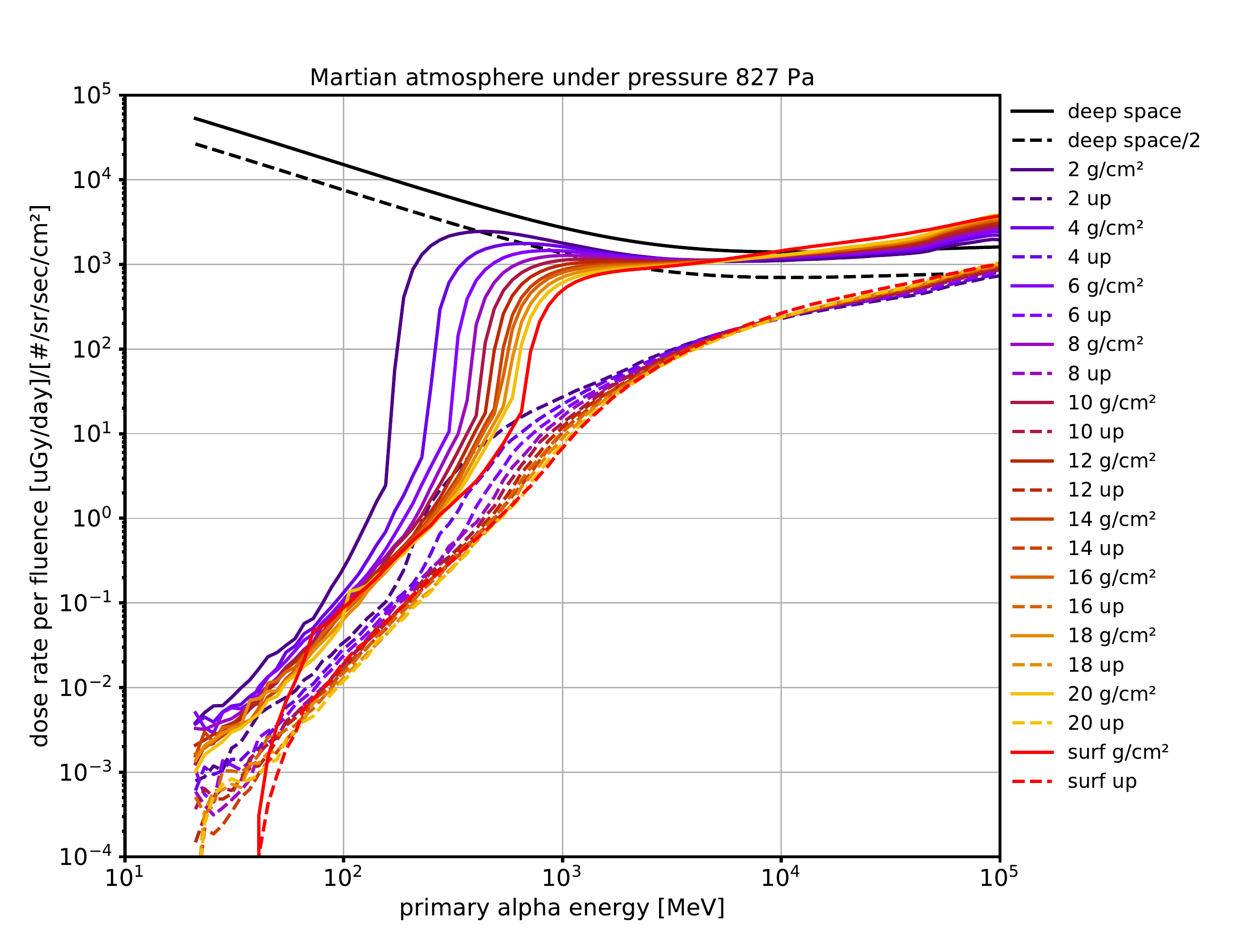}} 
	\end{tabular}
	\caption{Dose rate in water per primary flux function for protons (top) and $^4$He ions (bottom) at different depth (2,4,6...20 and surface) of the Martian atmosphere. The total vertical atmospheric depth under a surface pressure of 827 Pascal is about 23 g/cm$^2$. Solid lines are for the total dose rate while dashed lines represent the upward contribution. Black lines mark the deep space case similar to Figure \ref{fig:dosefunctions_surf}.  
	}\label{fig:dosefunctions_827Pa}
\end{figure}

On the surface of Mars, we calculated the dose rate per primary flux function for both primary protons and helium ions as shown in Figure \ref{fig:dosefunctions_surf}. 
For different surface pressure conditions which are different seasonal pressures measured at Gale Crater \citep{guo2015modeling}, these are ready-to-go dose functions versus the energy of primary particles reaching the top of the Martian atmosphere. 

As shown in the figure, there is a sharp decrease of surface dose contribution from primary protons with energies below the gray shaded area which is from 140 to 190 MeV/nuc. 
This 'cutoff' energy $E_c$ is equivalent to the energy required for a proton to penetrate through the total Martian atmospheric depth as shown in the proton-proton matrix in Figure \ref{fig:PLANETOMATRIX}a. 
Contribution to the surface dose below this primary energy is mainly from proton generated secondaries such as electrons, gammas, and neutrons such as illustrated in Figure \ref{fig:PLANETOMATRIX}b.

It is also shown that this cutoff energy depends on the surface pressure (proportional to the total atmospheric depth): higher surface pressure corresponds to a slightly larger cutoff energy as it is more difficult for particles to penetrate through the whole atmosphere. 
This is supported by observations \citep{rafkin2014} that the surface dose rate seen by MSL/RAD is anti-correlated with the surface pressure measured by the Rover Environmental Monitoring Station \citep[REMS,][]{gomez2012rems} on board MSL. 

The primary spectra used to fold with these functions are free space GCR or SEP spectra in units of particles/sr/cm$^2$/sec/MeV (which are first scaled by the energy bin width used for calculating these functions before the folding). 
And the total induced dose rate (in units of \textmu Gy/day) by certain given spectra is the integrated sum of the free space spectra multiplying the dose function. 
We also calculated and plotted the dose rate function induced by the primary protons and $^4$He ions in deep space with no planet or atmosphere shielding present and the full geometric angle is 4 $\pi$ as shown in the black solid lines in Figure \ref{fig:dosefunctions_surf}. 
The black dashed lines illustrate the scenario when only half of the full geometry should be considered such as near a planet or a large spacecraft which blocks about half of the isotropic free space flux (while no albedo secondaries are neglected). 
As shown, the dose function in deep space is rather similar to the ionization energy loss (e.g., the Bethe-Bloch function) process and it increases significantly for low-energy particles and flattens for relativistic high-energy ones. 

Colored dashed lines in Figure \ref{fig:dosefunctions_surf} represent the  contribution to dose by upward albedo secondaries. 
The contribution ratio is rather small, $<10\%$, especially at energies slightly above $E_c$ as the primary particles just had sufficient energy to make down to the surface and not sufficient remaining energy to knock out target nucleus in the regolith.
As primary particles have higher energies and thus higher remaining energies on the surface, they induce more spallation interactions with regolith to generate albedo secondaries and the upward dose contribution is slightly bigger.
At energies below the cutoff energy both downward and upward contributions to dose are very small and are from secondaries generated in the atmosphere (as few primaries could reach the ground) and the upward dose ratio is roughly constant at $\sim 20\%$. 

{In general, the upward dose, mostly produced in the
Martian regolith, is much lower than the downward dose, this suggests the effectiveness of shielding using the Martian regolith (e.g., inside a lava tube, near a cliff or underground) for potential human habitat on Mars. We also note that the generation of albedo particles may depend on the regolith composition. For instance, water content in the soil may thermalize fast neutrons \citep{boynton2012high} and change the albedo neutron spectra which will affect the upward dose. Although we do not expect any significant modifications of the total dose, detailed simulations with better knowledge of the Martian soil would be helpful for better quantifying the impact of the regolith composition on dose.}

We also note that there are some statistical uncertainties causing spikes in the functions especially at low energies where the statistics of secondaries generated in the atmosphere is relatively poor. 
Since their contribution to the total integrated dose is orders of magnitude lower than higher primary energies, this does not affect the calculations of the total dose. 
However, in the following Figures \ref{fig:dosefunctions_827Pa} and \ref{fig:doseEqfunctions_827Pa} where more functions are present, we smooth the dose curves to better illustrate the function dependence on atmospheric depth.

Figure \ref{fig:dosefunctions_827Pa} shows the same functions as in Figure \ref{fig:dosefunctions_surf} but for different atmospheric depth of 2,4,6...20 g/cm$^2$ down to the surface with a fixed surface pressure of 827 Pascals. The surface pressure is equivalent to a total column depth of about 23 g/cm$^2$. 
As clearly shown, at lower energies $< \sim 1$ GeV/nuc, the dose function per primary flux decreases as the atmospheric depth increases due to primary particles loosing energy as they transverse deeper in the atmosphere. 
For the same reason, the cutoff energy increases from $\sim$ 40 MeV at 2 g/cm$^2$ to $\sim$ 180 MeV at the surface. 
At energies higher than a few GeV, secondary generations are more frequent due to particle fragmentation and spallation interactions with the atmosphere and the dose is enhanced as the atmospheric is thicker. 
The upward dose also shows similar dependence on the atmospheric depth. 
At energies smaller than a few GeV, the atmospheric shielding effect against primary particles dominates and dose decreases as atmosphere is thicker. 
At higher energies, the upward dose is mainly contributed by secondaries generated in the regolith. It is higher at the surface and slightly decreases as the atmospheric altitude increases since upward particles loose their energies in the atmosphere as they transverse upwards. 

At high energies, the enhanced dose at thicker atmospheric depth is similar to the enhanced particle fluxes above the Pfortzer maximum on Earth \citep{pfotzer1936}. 
Since Earth has a global magnetosphere to efficiently shield against lower energy particles, this effect is also pronounced in the integrated dose/flux. 
At Mars which does not have a global magnetosphere, the atmospheric shielding against lower energy primaries is more dominating, especially during solar minimum when low-energy particles ($<$ GeV) are more abundant, and thus observations show an anti-correlation of the integrated surface dose rate versus the thickness of the atmosphere \citep{rafkin2014, guo2017dependence}.

\begin{figure}[ht!]
\centering
\begin{tabular}{cc}
\subfloat{ \includegraphics[trim=15 20 20 20,clip, scale=0.48]{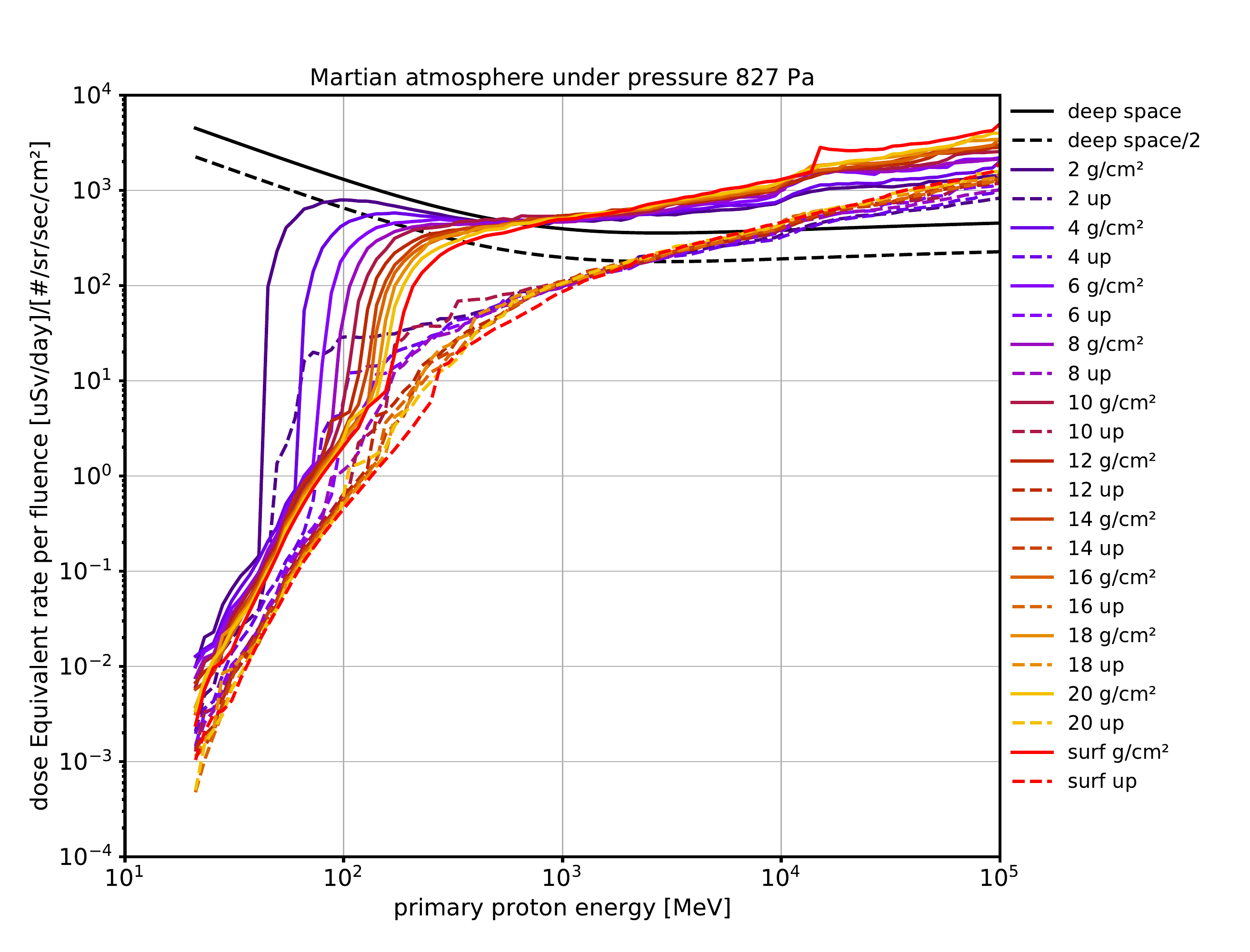}} \\
\subfloat{ \includegraphics[trim=15 20 20 20,clip, scale=0.48]{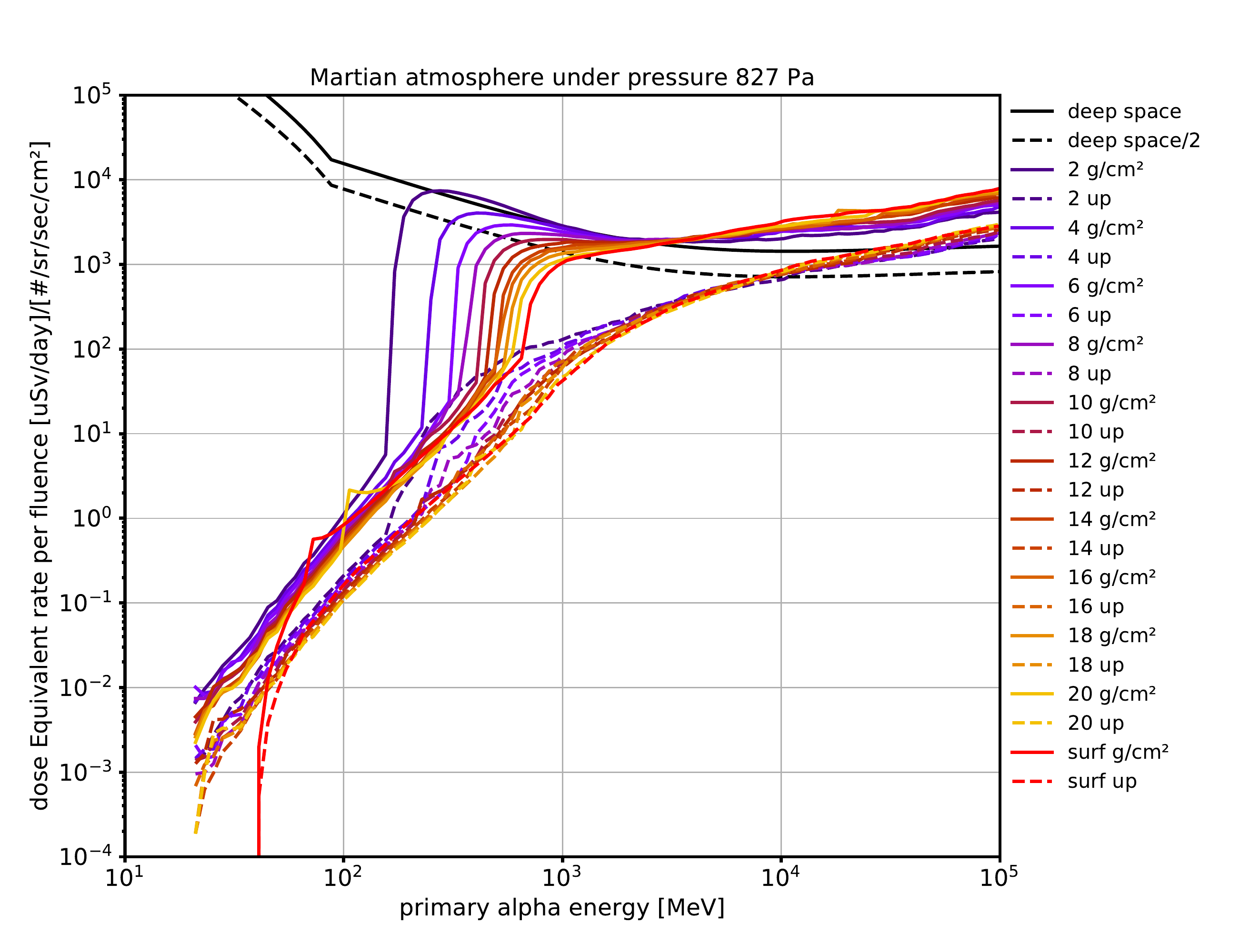}}
\end{tabular}
\caption{Dose equivalent rate per primary flux function for protons (top) and $^4$He ions (bottom) at different depth (2,4,6...20 and surface) of the Martian atmosphere with a surface pressure of 827 Pascal. Line colors and properties are set to be same as in Figure \ref{fig:dosefunctions_827Pa}.
}\label{fig:doseEqfunctions_827Pa}
\end{figure}

As explained in Section \ref{sec:functions}, we have also considered the dose equivalent rate functions which is an important factor 
and the results are shown in Figure \ref{fig:doseEqfunctions_827Pa}. 
The features of their dependence on energy and atmospheric depth are very similar to the dose functions. As $^4$He ions have higher charge and larger LET, their contribution to dose equivalent rate relative to the proton contribution is enhanced in comparison to the case of dose rate. 

It is important to note that the functions calculated and provided in this study should be used with the awareness of their assumptions and limitations. 
{First, the physics list used in the GEANT4 simulations are emstandard\_opt4 and QGSP\_BIC\_HP as described in Section \ref{sec:planetocosmics} and \citet{guo2018generalized}. 
Recently, we have compared four different physics lists in GEANT4 when applied to the Martian environment \citep{guo2019atris} and found that spallation processes can be more efficient in the Li\'ege Intra-nuclear Cascade (INCL) model than the Binary or Bertini cascade models. Thus the production of deuterium, tritium and $^3$He particles by primary protons and $^4$He particles could be better described by INCL in the energy range of $\sim$ 1-20 GeV. However, in the Martian environment, the contribution to dose and dose equivalent by such particles are significantly smaller than that by primary protons and $^4$He ions and other secondaries such as electrons and neutrons. 
Therefore the predictions of surface dose by different physics models are agreeing with each other reasonably well as shown by \citet{matthia2016martian}.} 

Moreover, only surface particles of proton, electron, position, $^4$He, $^3$He, deuteron, triton, neutron and gamma above 1 MeV are considered in the calculations of dose and dose equivalent. 
We did notice a slight increase in dose at lower primary energies ($< \sim$ 100 MeV) and high primary energies (larger than a few GeVs) when including more secondaries and having a lower energy cutoff in calculating dose. 
For instance, when including particles $<$ 1 MeV, the dose function increases by about 50\% at 25 MeV primary proton energy and about 15\% at 6400 MeV, mostly due to contributions by electrons, followed by positions. At 6400 MeV, the dose contributions from pions and muons have been estimated to  about 5\% which are not considered in the current dose functions. 
But the resulting total dose induced by an SEP or GCR spectra is not altered much. 
Based on a typical power-law SEP spectra with the energy-dependent flux of I(E)=I$_0$*(E/E$_0$)$^{-\gamma}$ where E$_0$ is 500 MeV, I$_0$ is 0.1/MeV/sec/sr/cm$^2$, and $\gamma$ is 2.5, we estimated the total dose rate also including secondary pions and muons and particles below 1 MeV.
The total dose rate is about 5\% larger than the results based on the dose functions provided here.
For GCR particles which have a much bigger high energy component, this difference is slightly bigger, that is, 15\% for dose rate and 25\% for dose equivalent rate. 
Besides, it is worth noting that when considering the full range of GCR particles, the proton and $^4$He ion combined contribution to the total dose (including also heavier GCR ions) is about 90\% and about 80\% to the total dose equivalent \citep{matthia2016martian}. 

\section{The RADMAREE Implementation}\label{sec:implementation}
\begin{figure}
    \centering
    \includegraphics[scale=0.6]{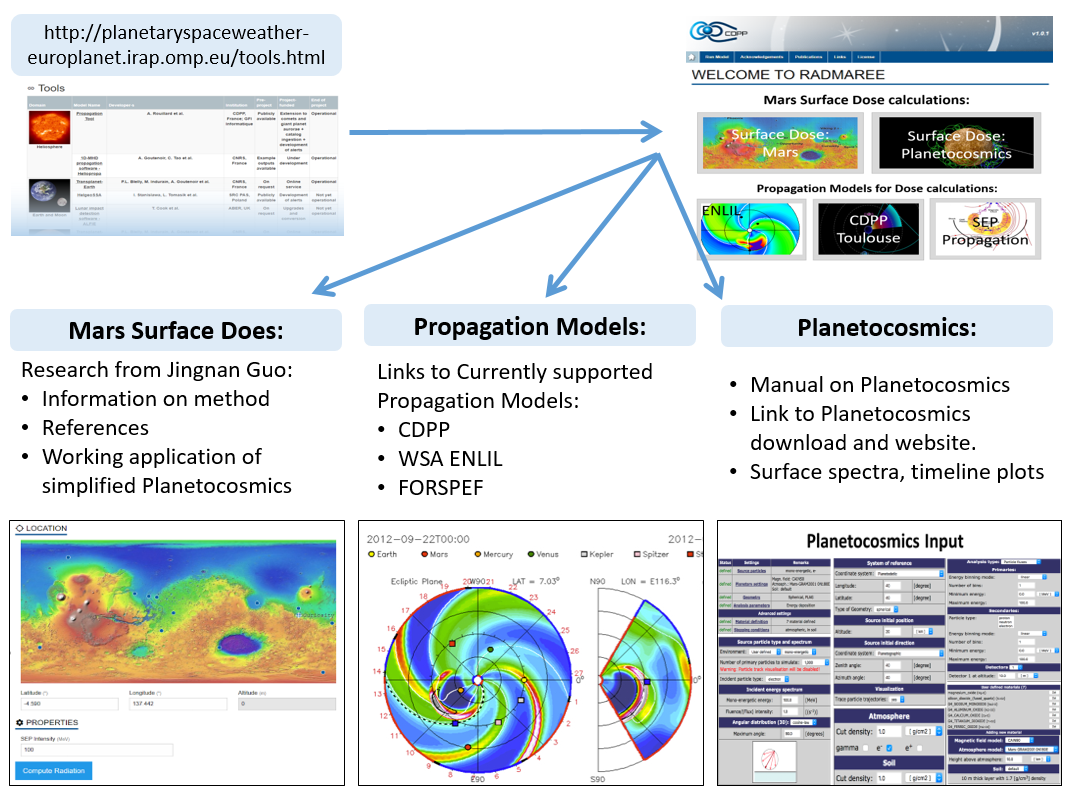}
    \caption{The online interface of the RADMAREE implementation (http://radmaree.irap.omp.eu/) as part of the European planetary space weather service. }
    \label{fig:RADMAREE}
\end{figure}

RADMAREE (RADiation MArtian Environment) is a simple to use application providing a quick and relatively accurate estimation of the radiation dose on the Martian surface. It is a facility supplied by CDPP and hosted at IRAP as supported by Europlanet RI at http://radmaree.irap.omp.eu/. As shown in Figure \ref{fig:RADMAREE}, RADMAREE is part of an integrated suite of tools developed under the Europlanet RI PSWS activity (http://planetaryspaceweather-europlanet.irap.omp.eu/tools.html ) where a number of other tools relevant for understanding and predicting the Martian space weather environment are also provided.

At the moment of writing this article, the service predicts the surface radiation dose due to a given SEP spectra with a power-law shape in the energy range of $\sim$ 100-800 MeV. This is based on the empirical linear correlation of the SEP intensity and surface dose as obtained by \citet{guo2018generalized} shown in Figure 7.  
With the implementation of the ready functions obtained in this work, a more accurate calculation of the surface dose can be achieved without requiring the energy range and shape of the input SEP spectra. In fact, these ready-to-go functions can be used to fold with any given SEP or GCR proton and helium ion spectra for obtaining the radiation dose on the surface of Mars and also at different depth of the atmosphere which may be relevant for different Martian seasons or different locations on the planet.

\section{Summary and Conclusions}
In order to prepare for future manned and robotic missions to Mars, especially during extreme and elevated conditions such as an SEP event, we have carried out detailed GEANT4 Monte Carlo simulations of particle interactions within the Martian atmospheric and regolith environment. 
We have calculated and obtained some ready-to-go functions which can be used to quickly convert any given SEP or GCR proton and helium ion spectra to the radiation dose on the surface of Mars and also at different depth of the atmosphere. 
These functions will be soon integrated into the {RADMAREE} under the Europlanet project which can be easily accessed by {the public as explained in Section \ref{sec:implementation}}. 

The dependence of these functions on the primary particle energy and the Martian atmospheric depth also reveals interesting physics of the particle interaction process with the Martian atmosphere and regolith.
Lower energy primary particles undergo strong shielding effect by the atmosphere while higher energy particles have more spallation and fragmentation interactions generating secondaries in the atmosphere at the surface. 
Therefore with the Martian atmosphere, the radiation dose resulted from all secondaries induced by such a high energy primary particle may exceed the deep space dose of this primary particle.
This highlights the importance of better detecting and modeling the high energy component (larger than $\sim$ GeV/nuc) in the GCR and SEP spectra for mitigating the Martian surface radiation environment with optimized shielding in the future exploration missions. 

\begin{acknowledgements}
The work is supported by DLR and DLR's Space Administration grant numbers 50QM0501, 50QM1201 and 50QM1701 to the Christian Albrechts University, Kiel.
M.G., D.M. and Z.L.P. are supported by Europlanet RI which also supported some travel for J.G.
R.F.W.S., J.G. and M.G. acknowledge the International Space Science Institute, which made part of the collaborations in this paper through the ISSI International Team 353 “Radiation Interactions at Planetary Bodies”.
The authors and the editor thank Fabiana Da Pieve and an anonymous referee for their assistance in evaluating this paper.
\end{acknowledgements}

\bibliography{msl_rad_guo}
\Online
\begin{appendix}
\textbf{{Appendix: Calculation of dose and dose equivalent \\}}
When calculating doses in a given volume for a certain radiation environment it is necessary to differentiate between the physical quantity absorbed dose D (see Eqs. \ref{eq:dose_dn} and \ref{eq:dose_up}) and the biologically weighted dose equivalent which is Q $\cdot$ D where Q is the quality factor which is introduced to take into account the biological effectiveness of different types of radiation \citep{icrp103}. The quality factor is defined as a function of the unrestricted linear energy transfer (LET) which is the mean energy loss by certain type of charged particles due to electronic interactions per path length. The unrestricted linear energy transfer is identical to the electronic stopping power and can be calculated, under certain conditions, from the Bethe-Bloch equation \citep{bethe1932bremsformel}. 

In a perfect secondary particle equilibrium the absorbed dose rate can be calculated from the stopping power (or equivalently the unrestricted linear energy transfer) as $D=\int dE F(E) \cdot S(E)/\rho$ where E is energy, F(E) is particle fluence (as shown in Eq. \ref{eq:matrix_multiply}) and S(E) and $\rho$ are electronic stopping power and density in the volume of interest, for instance a detector or a human organ. 
In a realistic scenario, however, this approximation has to be considered with caution as it relies on several assumptions that may be violated: the stopping power has to be fairly constant over the sensitive volume, and the particle creation and loss rate within the volume has to be identical (or very similar) to the surroundings (secondary particle equilibrium). 
While these assumptions are generally well fulfilled for the strongly penetrating radiation field of galactic cosmic radiation, this is not necessarily the case for solar particle events which may contain a substantial amount of lower energetic particles especially in deep space. On the surface of Mars, the ratio of the low energy SEPs are considerably smaller due to the effective shielding of the Martian atmosphere against lower energy SEPs. 

When considering a radiation field with significant amount of low-energy particles, the geometry of the sensitive volumes is important. 
Specifically speaking, a particle with low energy would slow down significantly or even lose all its energy within a small layer at the surface of the volume of interest and cannot reach deeper lying areas, e.g., inner organs in a human body. 
In such a case the deposited energy in the volume and the corresponding absorbed dose can be estimated with specific calculations taking into account of the geometry and material of the concerned volume. 
In these simulations the energy deposition for a given particle can be calculated in steps along its track (and multiplied by the corresponding quality factor if the dose equivalent is calculated) and the dose is the accumulated energy deposition divided by the mass within the volume. 
\end{appendix}


\end{document}